\documentclass[11pt]{article}
\usepackage{times}
\usepackage{sidecap}
\usepackage{geometry}
\usepackage{floatrow}
\usepackage{float}
\usepackage{hyperref}
\usepackage{graphics,graphicx}
\usepackage{subcaption}
\usepackage{epsf,wrapfig}
\usepackage{ulem} 

\usepackage{natbib,bm}

\usepackage{paralist}
\let\olditem\item
\renewenvironment{thebibliography}[1]{%
  \section*{\refname}
  \let\par\relax
  \renewcommand{\item}[1][]{\olditem[\textbullet]}%
  \inparaenum}{\endinparaenum}

\newcounter{aff}

\DeclareGraphicsExtensions{.jpg,.pdf,.png,.eps}

\usepackage{caption}
\usepackage{xcolor}

\usepackage{tcolorbox}

\bibpunct{(}{)}{;}{a}{}{,}

\usepackage{soul}
\usepackage{xcolor}

\definecolor{darkcyan}{rgb}{0.0, 0.0, 0.7}

\definecolor{boxcolor}{RGB}{193, 229, 255}
\definecolor{figboxcolor}{RGB}{193, 229, 255}
\definecolor{boxcolorgray}{RGB}{227, 227, 227}
\definecolor{darkcyan}{rgb}{0.0, 0.0, 0.7}

\sethlcolor{boxcolorgray}

\def\gtrsim{\lower 2pt \hbox{$\, \buildrel {\scriptstyle >}\over
{\scriptstyle \sim}\,$}}
\def\lesssim{\lower 2pt \hbox{$\, \buildrel {\scriptstyle <}\over
{\scriptstyle \sim}\,$}}

\def\approxlt{\lower.2em\hbox{$\buildrel < \over \sim$}}
\def\approxgt{\lower.2em\hbox{$\buildrel > \over \sim$}}

\def\be{\begin{equation}}
\def\ee{\end{equation}}
\def\bea{\begin{eqnarray*}}
\def\eea{\end{eqnarray*}}

\def\jref@jnl#1{{\rm#1}}
\def\aj{\jref@jnl{AJ}}			
\def\araa{\jref@jnl{ARA\&A}}		
\def\apj{\jref@jnl{ApJ}}			
\def\apjl{\jref@jnl{ApJ}}		
\def\apjs{\jref@jnl{ApJS}}		
\def\ao{\jref@jnl{Appl.~Opt.}}		
\def\apss{\jref@jnl{Ap\&SS}}		
\def\aap{\jref@jnl{A\&A}}		
\def\aapr{\jref@jnl{A\&A~Rev.}}		
\def\aaps{\jref@jnl{A\&AS}}		
\def\azh{\jref@jnl{AZh}}			
\def\baas{\jref@jnl{BAAS}}		
\def\jrasc{\jref@jnl{JRASC}}		
\def\memras{\jref@jnl{MmRAS}}		
\def\mnras{\jref@jnl{MNRAS}}		
\def\pra{\jref@jnl{Phys.~Rev.~A}}	
\def\prb{\jref@jnl{Phys.~Rev.~B}}	
\def\prc{\jref@jnl{Phys.~Rev.~C}}	
\def\prd{\jref@jnl{Phys.~Rev.~D}}	
\def\pre{\jref@jnl{Phys.~Rev.~E}}	
\def\prl{\jref@jnl{Phys.~Rev.~Lett.}}	
\def\pasp{\jref@jnl{PASP}}		
\def\pasj{\jref@jnl{PASJ}}		
\def\qjras{\jref@jnl{QJRAS}}		
\def\skytel{\jref@jnl{S\&T}}		
\def\solphys{\jref@jnl{Sol.~Phys.}}	
\def\sovast{\jref@jnl{Soviet~Ast.}}	
\def\ssr{\jref@jnl{Space~Sci.~Rev.}}	
\def\zap{\jref@jnl{ZAp}}			
\def\nat{\jref@jnl{Nature}}		
\def\iaucirc{\jref@jnl{IAU~Circ.}}
\def\aplett{\jref@jnl{Astrophys.~Lett.}}
\def\apspr{\jref@jnl{Astrophys.~Space~Phys.~Res.}}
\def\bain{\jref@jnl{Bull.~Astron.~Inst.~Netherlands}}
\def\fcp{\jref@jnl{Fund.~Cosmic~Phys.}}
\def\gca{\jref@jnl{Geochim.~Cosmochim.~Acta}}
\def\grl{\jref@jnl{Geophys.~Res.~Lett.}}
\def\jcp{\jref@jnl{J.~Chem.~Phys.}}	
\def\jgr{\jref@jnl{J.~Geophys.~Res.}}	
\def\jqsrt{\jref@jnl{J.~Quant.~Spec.~Radiat.~Transf.}}
\def\memsai{\jref@jnl{Mem.~Soc.~Astron.~Italiana}}
\def\nphysa{\jref@jnl{Nucl.~Phys.~A}}
\def\physrep{\jref@jnl{Phys.~Rep.}}
\def\physscr{\jref@jnl{Phys.~Scr}}
\def\planss{\jref@jnl{Planet.~Space~Sci.}}
\def\procspie{\jref@jnl{Proc.~SPIE}}

\def\ion#1#2{#1$\;${\small\rm\@Roman{#2}}\relax}

\newcount\lecurrentfam
\def\LaTeX{\lecurrentfam=\the\fam \leavevmode L\raise.42ex
\hbox{$\fam\lecurrentfam\scriptstyle\kern-.3em A$}\kern-.15em\TeX}
\def\sizrpt{
(\fontname\the\font): em=\the\fontdimen6\font, ex=\the\fontdimen5\font
\typeout{
(\fontname\the\font): em=\the\fontdimen6\font, ex=\the\fontdimen5\font
}}
\definecolor{dark}{rgb}{ 0.34, 0.09, 0.49}

\usepackage{fancyhdr,url}
\usepackage{transparent}
\usepackage{tablefootnote}
\newcommand{\sgra}{Sgr~A*}
\usepackage{mathpazo}

\begin{document}

\noindent
\begin{raggedright}
{\Huge\bfseries Probing fundamental physics with ALMA2040 mm-wavelength observations of neutron stars \\[0.5em]
\Large 
}\end{raggedright}

\noindent

Kuo Liu$^{1,2}$, Marilyn Cruces$^{3,4,2}$, Gregory Desvignes$^2$, Michael Kramer$^2$, Mariafelicia De Laurentis$^5$, Natalia Lewandowska$^6$, Nutnicha Masoongnoen$^7$, Dieu D. Nguyen$^8$, Lijing Shao$^9$, Pablo Torne$^{10,2}$, Jompoj Wongphechauxsorn$^{11}$

\vspace{1em}
{
\footnotesize{
\noindent $^1$ State Key Laboratory of Radio Astronomy and Technology, Shanghai Astronomical Observatory, CAS, 80 Nandan Road, Shanghai 200030, P. R. China \\
$^2$ Max-Planck-Institut für Radioastronomie, Auf dem Hügel 69, D-53121 Bonn, Germany \\
$^3$ Centre of Astro-Engineering, Pontificia Universidad Católica de Chile, Av. Vicuña Mackenna 4860, Santiago, Chile \\
$^4$ Department of Electrical Engineering, Pontificia Universidad Católica de Chile, Av. Vicuña Mackenna 4860, Santiago, Chile \\
$^5$ Dipartimento di Fisica “E. Pancini”, Universit`a di Napoli “Federico II”, Compl. Univ. di Monte S. Angelo, Edificio G, Via Cinthia, I-80126, Napoli, Italy \\
$^6$ Department of Physics and Astronomy, State University of New York at Oswego, Oswego, NY 13126, USA \\
$^7$ National Astronomical Research Institute of Thailand (Public Organization), Chiangmai, 50180, Thailand \\
$^8$ Department of Astronomy, University of Michigan, 1085 South University Avenue, Ann Arbor, MI 48109, USA \\
$^9$ Kavli Institute for Astronomy and Astrophysics, Peking University, Beijing 100871, China \\
$^{10}$ Institut de Radioastronomie Millim\'etrique (IRAM), Avenida Divina Pastora 7, Local 20, E-18012, Granada, Spain \\
$^{11}$ Julius-Maximilians-Universität Würzburg, Institut für Theoretische Physik und Astrophysik, Lehrstuhl für Astronomie, Emil-Fischer-Straße 31, D-97074 Würzburg, Germany \\
}
}
\begin{tcolorbox}[boxsep=3pt,left=0pt,right=0pt,top=1pt,bottom=1pt,colframe=boxcolor,colback=boxcolor]
\justify
\textbf{Abstract:} 
Bursty or (rotation-induced) pulsed coherent radio emission from pulsars, magnetars, and fast radio transients originates in some of the most extreme environments in the Universe. However, the exact physical mechanism behind the scenes remains poorly understood. ALMA2040 will allow for systematic exploration of the millimeter-wavelength spectral window for pulsars and fast radio transients for the very first time, offering unprecedented opportunities to discover new observational phenomena and advance our understanding of emission physics. Beyond this, ALMA2040 provides a unique means to unveil the neutron-star population in the Galactic Centre (GC), enable novel tests of fundamental physics in the vicinity of the supermassive black hole \sgra, probe the potential existence of dark matter in the GC, and investigate the dense, highly dynamic interstellar and magnetic environment of this region.
\end{tcolorbox}

\newpage{}

\vspace{-0.7cm}
\section{Scientific context and open questions} \label{sec:science}
When observed at radio frequencies, the sky is largely populated by transient events, with timescales as short as a fraction of a millisecond. Among such short-lived transients, the coherent and most energetic ones at microsecond to millisecond timescales come from pulsars and fast transients. Pulsars are rotating neutron stars emitting beams of electromagnetic radiation out of its magnetic poles across the full electromagnetic spectrum. Magnetars are a rare sub-class of pulsars known to host the strongest magnetic fields in the Universe reaching up to $10^{14}$--$10^{15}$\,G. As time-domain radio surveys continue to improve in sensitivity and techniques, new types of radio transient phenomena emerge. Highlights are fast radio bursts \citep[FRBs;][]{lbm2007}, bright flashes of extragalactic origin, and more recently ultra-long period transients \citep[ULPTs;][]{hrm2023} which show periodicities of up to hours and pulse durations up to a few minutes \citep[for a review, see][]{rhc26_review}. Pulsars and fast transients are unique laboratories that offer extreme physical conditions such as extremely high plasma densities and magnetic fields. Pulsars display incredibly regular rotational periods and long-term stability that can reach that of atomic clocks, which make them highly sensitive probes into spacetime properties. Similarly, the immense amount of energy available to power a pulse that may be visible at large distances (half way the Universe in the case of FRBs), makes them powerful tools for cosmology. 

However, the emission mechanism of pulsars remains one of the most enduring puzzles in astrophysics. Open questions such as the trigger of coherent radio emission and the exact location of emission region in the pulsar magnetosphere have yet to be fully answered. The physical mechanism that produces FRBs is even further from being fully understood and raises the fundamental question: What powers transient radio emission across the Universe? Is it inherently tied to neutron stars with different energy budgets, or a more universal mechanism operating in diverse compact objects at different scales? As for ULPTs, the nature of the progenitor is also unclear. If they are neutron stars, they would lie deep in the so-called death valley where, based on standard pulsar emission models, coherent radio emission should no longer be possible due to insufficient pair production. Alternatively, ULPTs might represent the first radio-emitting white dwarfs detected via periodic pulses. 

In radio, most pulsar and fast transient studies have so far concentrated at decimeter wavelengths ($\sim$1\,GHz) which optimize the observing sensitivity. Consequently, the millimeter (mm) wavelength regime of their spectrum has barely been explored. To date, out of the 5000+ pulsars and FRBs known, only five pulsars and three magnetars have reported detections at frequencies above 40\,GHz \citep{kjdw97,mkt+97,tek+15,lyw+19,tbb+22}, and no FRB has been detected above 8\,GHz \citep{gajjar18}. This is likely due to a combination of instrinsic unpredictability of their emission and the lack of instruments with prompt response to follow-up new FRB discoveries. Altogether, this makes the mm-wavelength window a ``hidden gem'' in pulsar and FRB studies, which however can only be effectively explored with ALMA2040.

\subsection*{Shaping coherent radio emission from neutron stars over a 100-GHz scale} \label{ssec:spec}
\vspace{-0.1cm}

\begin{wrapfigure}[22]{r}{0.49\textwidth}
    \centering
    \vspace{-40pt}
    \includegraphics[width=\linewidth]{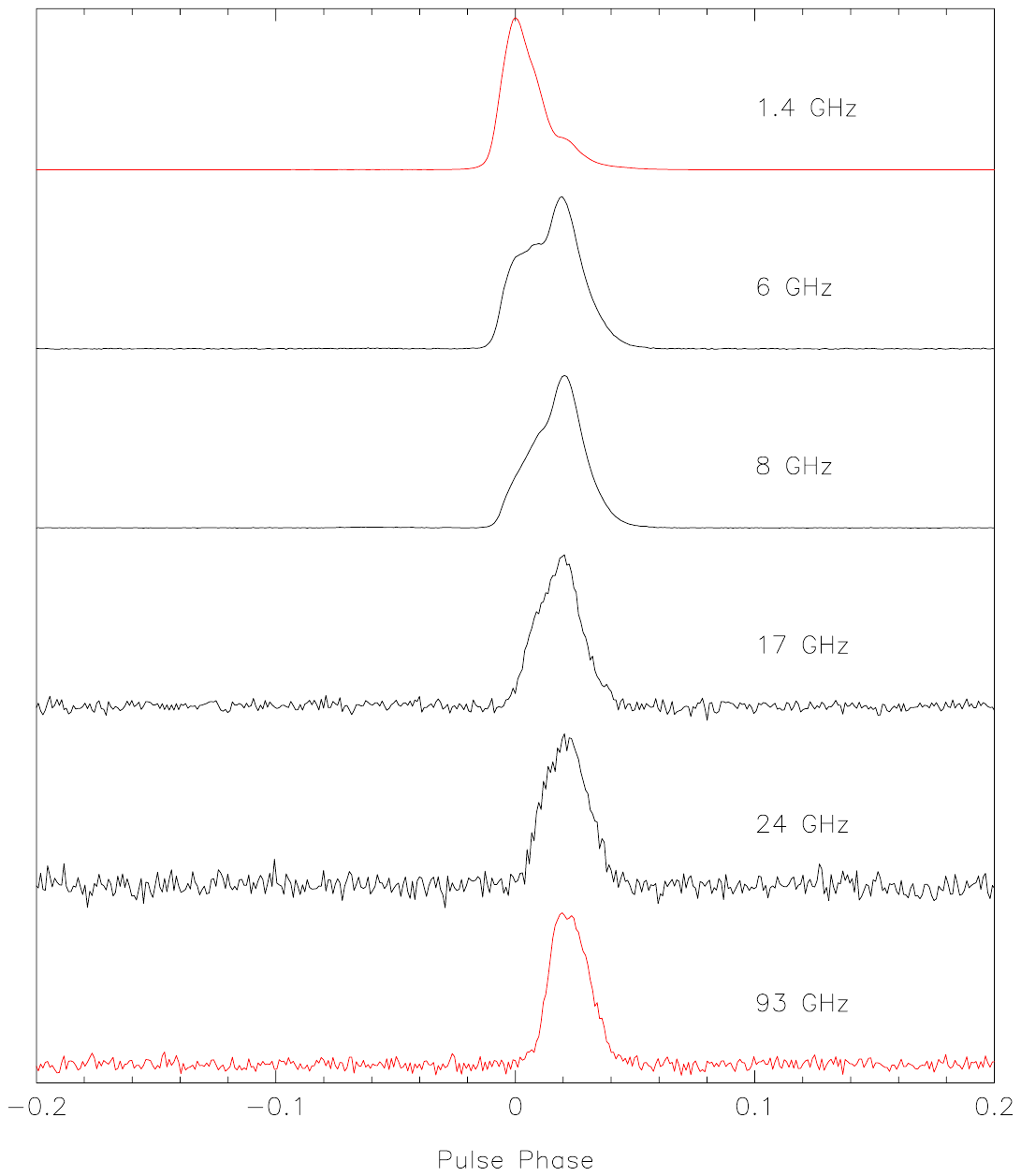}
    \vspace{-70pt}
    \caption{Pulse profile of the Vela pulsar as a function of frequency, covering a frequency range of $\sim100$\,GHz. This figure is reproduced from \citet{lyw+19}.\label{fig:profiles}}
\end{wrapfigure}

The investigation into the mm-wavelength pulsar spectrum offers a unique handle to search for new emission mechanisms in between the radio and infrared band which are predicted by a group of theories. While coherent emission is required at low radio frequencies, some models include an incoherent emission component that dominates at high radio frequencies \citep{mic82,Crusius-Watzel2001}. Inverse Compton scattering of low-frequency radio photons is also predicted to introduce specific emission components rising somewhere in between the radio and infrared bands \citep{Lyutikov2013}. Meanwhile, emission mechanisms involving pair plasma discharges predict a cutoff in the spectrum at high radio frequencies ($\gtrsim$50\,GHz) \citep{Philippov2020}. The complexity in pulsar's radio spectrum at high frequencies, e.g., frequency-dependent emission components (see Figure~\ref{fig:profiles}), has been seen in a few cases \citep{hje15}. However, a generic conclusion needs to be drawn from a large sample study, which is only possible with ALMA2040. Simulation suggests that the sensitivity of ALMA2040 will allow to detect (with a significance of 10$\sigma$) a quarter of the currently known radio pulsars at a 100-GHz frequency, including 15\% of the millisecond pulsar populations. This will facilitate at least a factor of three of what is possible with the current ALMA sensitivity. 

Pulsars exhibit frequency-dependent pulse profiles that have been generally associated with the magnetic-field geometry where the emission is generated. Based on the traditional ``radius-to-frequency mapping'' (RFM) relation in pulsar emission geometry, emission at high radio frequencies originates from the deepest (and narrowest) region in the pulsar magnetosphere, i.e., the so-called ``polar cap'' region that is close to the pulsar surface at its magnetic poles. By combining the mm-wavelength pulse profiles with those at lower frequencies and modelling their shapes all together, a 3-D mapping of the pulsar magnetosphere can be constructed from a few tens of NS radius to the light cylinder \citep[e.g.,][]{naa+11}. This will provide novel insight into our understanding of pulsar emission geometry.

\subsection*{The origin of fast radio transients}
\vspace{-0.1cm}

By far, the strongest clue to the nature of FRB's progenitor comes from the detection in 2020 of a burst from the Galactic magnetar, SGR 1935+2154. Such pulse displayed FRB-like features although still a few orders of magnitude less energetic than average FRBs \citep{CHIME2020_burst}. Phenomenologically, FRBs are classified into (apparent) one-offs and repeaters. From repeaters, through variations in the bursts' in dispersion and rotation measure, we have learned that FRBs can originate in complex magneto-ionic environments \citep{ACS2023}. This type of revenue can be offered by the environment in the Galactic Centre \citep[GC;][]{Desvignes2018,vcl2025}. 

\begin{wrapfigure}[25]{r}{0.49\textwidth}
    \centering
   \vspace{-20pt}
    \includegraphics[width=\textwidth]{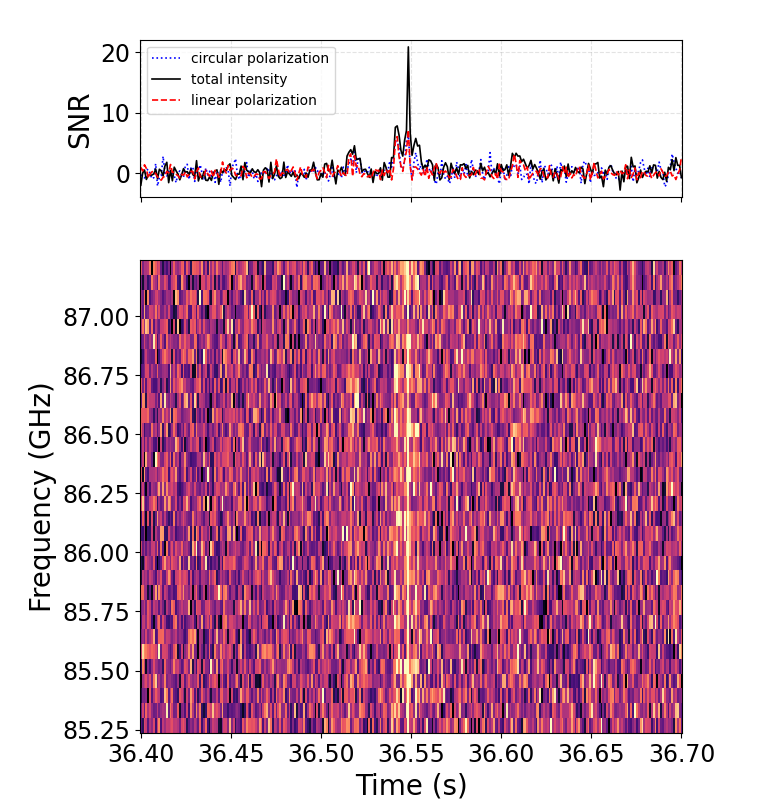}
    \caption{Single pulse from the GC magnetar, PSR J1745$-$2900, at 86 GHz with ALMA's phased array mode. Top panel: The pulse in total intensity (black), linear (red) and circular (blue) polarization. Bottom panel: Dynamic spectra along the 2-GHz bandwidth.}
    \label{fig:pulse}
\end{wrapfigure}

Interestingly, several magnetars have shown that coherent radio emission can extend into the mm regime, up to $\sim$350 GHz and may even exhibit flat or inverted spectral indices \citep{tde+17,tbb+22}, contrary to the standard steep pulsar spectral index. If a fraction of FRBs originate from similar magnetars, then a subpopulation could be detectable as well at mm-wavelength. High-frequency detections are particularly valuable since they provide cleaner access to intrinsic burst widths, microstructure, and signatures of the immediate environment that are diluted at GHz frequencies (See Fig.~\ref{fig:pulse}). This is because propagation effects become negligible at mm-wavelength. 

The most realistic strategy to follow-up FRBs with ALMA is likely targeting repeating FRBs. Among those, the seemingly periodically active repeaters \citep{Chime+2020,Cruces2020} are of particular interest given tens to hundreds of bursts detected per hour at peak activity phases \citep{bcc+25,ecc2025}. With its current PAM configuration with 2-GHz effective bandwidth, ALMA reaches burst fluence thresholds of 0.07--0.2\,Jy~ms for millisecond-duration bursts \citep{vcl2025}. A system with 10x better sensitivity would extend the detection horizon for typical FRB energies of $10^{40}$ erg from $z \sim 1$ to $z \sim 3-6$, making mm-wavelength detections of cosmological FRBs feasible. For ULPTs, although they are bright Galactic sources \citep[fluence $\sim 100$ Jy\,s;][]{hrm2023}, all detections so far lie below the GHz range. If their progenitors had a flat spectral index, as for magnetars, they could be detectable up to mm wavelengths. However, if their emission is pulsar-like, the steep spectral index would likely place them below ALMA’s current sensitivity. A boost in sensitivity (e.g., a factor of several) of mm observations would be required in this case.

\subsection*{Studying pulsars and transients in the Galactic Centre} \label{ssec:GC}
\vspace{-0.1cm}

The GC provides a unique environment for pulsar and transient studies. Discovering such sources in the GC would allow for a range of highly important and unique astrophysical investigations, with profound implications for fundamental physics.

\underline{Testing gravity in the strong-field regime:} The presence of the supermassive black hole, \sgra\, makes the GC an exceptional laboratory for gravity experiment \citep{GRAVITYCollaboration2019,EHT2022}. In particular, finding a pulsar in a sufficiently tight orbit around the \sgra\ would allow for precision timing experiment of the pulsar to monitor its orbital motion. Pulsar timing with ALMA2040 at mm-wavelengths is uniquely suited to this task, as such observations preserve the sharp pulse features and arrival-time precision on which these tests depend. This could yield measurements of the black hole's mass, spin, and in the most favorable cases, its quadrupole moment. Such measurements would enable precision tests of fundamental principles in our black-hole physics, including the cosmic censorship conjecture and the no-hair theorem \citep{lwk12,pwk16,hs2024}. 

\underline{Probing the dark-matter environment in the GC:} A compelling dark matter candidate is axion, a hypothetical pseudoscalar boson originally proposed to resolve the strong charge--parity problem in quantum chromodynamics \citep{PQ1997, weinberg1978new, wilczek1978}. In the presence of a strong magnetic field, axions can convert into photons and provide a promising observational signal \citep{Raffelt2007}. The GC magnetar, PSR J1745$-$2900, offers an ideal environment for searching for axion-photon conversion signals \citep[e.g.][]{darling2020search}. Observationally, ALMA band 4 has placed upper limits on the axion-photon coupling of $g_\gamma \gtrsim 2\times10^{-11}$ GeV$^{-1}$ and $g_\gamma \gtrsim 2\times10^{-13}$ GeV$^{-1}$, depending on the assumed dark matter density profile \citep{DeMiguel2026}. With the current ALMA, detection of axion-induced signals from PSR J1745$-$2900's magnetosphere across most frequency bands requires 100-hour observing time \citep{Masoongnoen2026}, with the caveat of assuming that magnetar's magnetosphere is stable over time \citep{tde+17}. With ALMA2040, the required observing time becomes an order of magnitude shorter and thus feasible, which would lead to a significantly tighter constraint on the axion-photon coupling \citep{Masoongnoen2026}. Moreover, if a binary pulsar is found in the GC, the distribution and properties of dark matter in the region would influence pulsar's orbital dynamics and imprint characteristic features on the timing signal, which in turn could probe the small-scale structure of dark matter \citep{hsz23}, as well as gravitational-atom and soliton profiles \citep{ychs25}.

\underline{Studying fast transients from the GC environment:} The GC can also play a key role in the study of fast transients. Given the highly magnetised and dynamic environment found to be associated with many repeating FRBs, it is natural to link to similar conditions observed in the central parsec of the GC \citep{Desvignes2018}. This makes the GC a promising site for probing FRB-like activity within our own Galaxy \citep{vcl2025}. The GC magnetar offers a ready-made venue to study FRB-like busty emissions in this highly magnetized environment, an ideal opportunity to investigate the magnetar hypothesis for the origin of FRB. 

\begin{wrapfigure}{r}{0.65\textwidth}
    \centering
    \includegraphics[width=\textwidth]{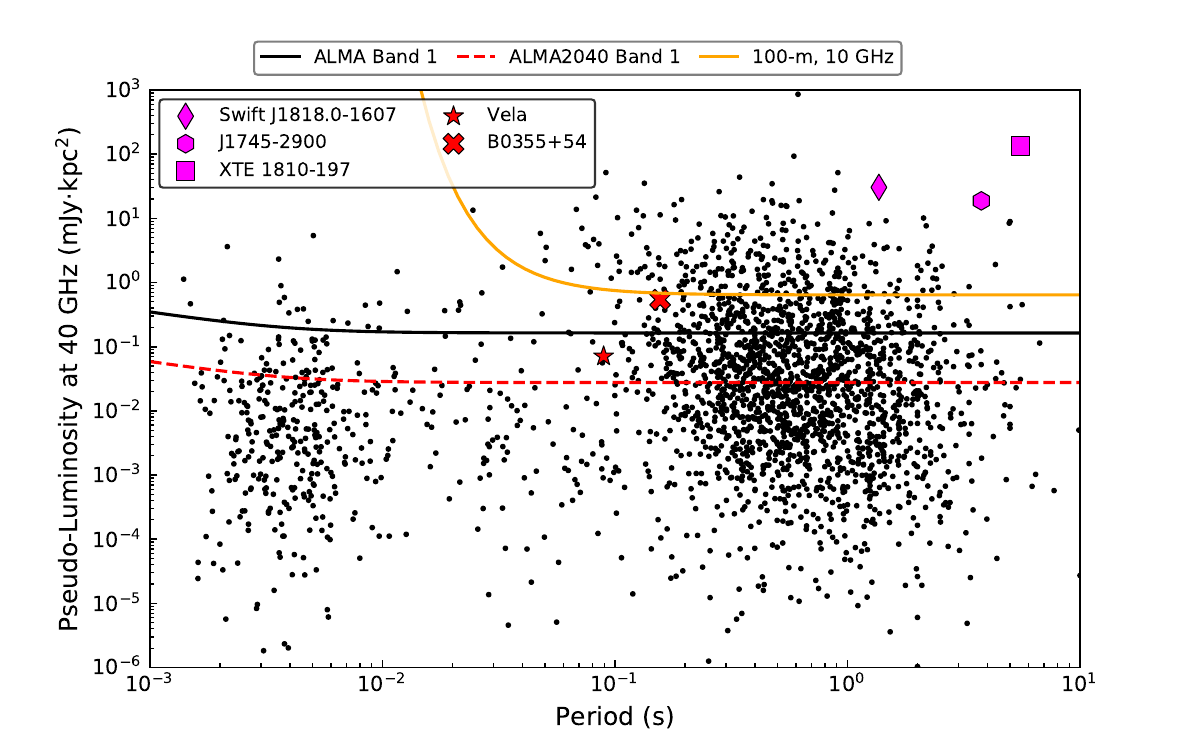}
    \caption{Sensitivity curves of GC pulsar surveys, with the current ALMA Band 1, ALMA2040 Band 1 and a 100-m class telescope. The assumed integration time is 3\,hr and the detection threshold is 10$\sigma$. The phasing efficiency was assumed to be 0.85 which also accounts for signal loss due to low-bit quantisation. The duty cycle of the signal is assumed to be 5\% based on available observations by far \citep{lyw+19}. The 10-GHz search sensitivity was converted to the 40-GHz pseudo luminosity by assuming a spectral index of $-1.6$. The black points represent a simulated pulsar population in the GC, created by following the procedures detailed in \citet{lde+21}. For pulsars with measured spectral index, we used that to extrapolate the flux density to 40\,GHz, assuming a power-law spectrum. Otherwise, we employed a random value drawn from a normal distribution, with mean of $-1.6$ and standard deviation of 0.54 as reported in \citet{jvk+18}.}
    \label{fig:gc}
\end{wrapfigure}

\underline{Uncovering the pulsar population in the GC:} The GC is rich in massive stars and is therefore expected to host a large population of neutron stars. However, only one radio pulsar has so far been found within a parsec of \sgra\, and none in a close orbit around it. This scarcity of pulsar discoveries has been referred to as ``the missing pulsar problem'' \citep{do14}. Nevertheless, there are clear observational obstacles to pulsar searching in the GC. One of the greatest challenges for low frequency ($\sim1\,$GHz) searches is interstellar scattering, which broadens the width of the pulsar signal and largely reduces its detectability. In the strong-scattering scenario, frequencies $\gtrsim15$\,GHz are anticipated to be necessary for discovering normal pulsars, and $\gtrsim 40~{\rm GHz}$ would be required for millisecond pulsars \citep{mkf+10,mk15}. On the other hand, pulsars typically exhibit a steep spectrum ($S_{\nu} \propto \nu^{\alpha}$, $\alpha \sim -1.6$) and are therefore faint at high frequencies \citep{jvk+18}. The primary limitation currently preventing new discoveries at mm-wavelengths is insufficient sensitivity \citep{lde+21,tde+21,tle+23}. Therefore, ALMA2040 is crucial for the task of detecting pulsars and fast transients in the GC. Given the synthetic beam size of phased ALMA, ALMA searches will focus on the very central region of the GC, i.e., within approximately 10\,arcsec around the \sgra. Figure~\ref{fig:gc} provides results of simulated pulsar surveys (for periodic signals) toward the GC. With a 3-hr-long integration at 40\,GHz with ALMA2040, we expect to find $\sim$45\% of the pulsars, including $\sim$20\% of the millisecond pulsars that could exist in the GC. 

\section{ALMA2040: Technical Requirements}

The anticipated flux density of pulsars and fast transients in the ALMA band is low. Extrapolating from low-frequency observations, for pulsars it should be mostly below 10\,$\mu$Jy \citep[e.g.,][]{kjdw97,mkt+97,lyw+19}. Therefore, an achievable sensitivity of $\sim$1\,$\mu$Jy with reasonable integration time (e.g., 1\,hr) is required to yield a significant detection (e.g., 10$\sigma$) of these sources. An improvement of approximately an order of magnitude over the current ALMA sensitivity, which can be achieved through a combination of enhanced line sensitivity and the increased instantaneous bandwidth, would satisfy this requirement. This major sensitivity enhancement is the key instrumental distinction between ALMA2040 and the Wideband Sensitivity Upgrade of ALMA2030 for pulsar and fast transient studies. 

Pulsar observations primarily employ the phased-array mode (PAM) of ALMA. The current PAM output data format would suffice for future observations: a time resolution down to 8\,$\mu$s, with full Stokes information and a modest frequency resolution of 62.5\,MHz, using all available spectral windows in each frequency band. Polarization capacity is highly desired, as most pulsars and magnetars often exhibit significant polarization fractions, reaching up to nearly 100\%. Regarding data dynamic range, 2-bit resolution in the phased-array voltage data is likely sufficient, since the strength of the incoming signal is minimal compared to the system temperature.  

Additionally, we anticipate that ALMA2040 will offer PAM with multiband capacity and in rapid-response observations. Multiband capacity spanning a frequency range of e.g., at least 512\,GHz instantaneously within the 35--1000-GHz window, is crucial for simultaneously covering a wide range in the spectrum of pulsars, magnetars and fast transients. This is particularly essential for FRB and magnetar studies, because it is the only way to study their true spectra which are time-variable. For pulsar studies, it will greatly improve observing efficiency and allow for studies with much larger samples. Instantaneous bandwidth in each band, especially in the lowest-frequency band, will also help improve the sensitivity for pulsar searching in the GC. The ability to conduct rapid-response observations in PAM is critical for promptly following up transient sources and enabling synergistic studies with other facilities.

\subsection{Simulations}

In the simulations carried out in Section~\ref{sec:science}, we adopted the compact array configuration provided by the simulation tool GHOST\footnote{\url{https://alma-ghost.jb.man.ac.uk}}. The assumed sensitivity of ALMA2040 in the simulations was 0.5\,$\mu$Jy for 1-hr integration time, approximately 10 times the current ALMA. For ALMA2040, we anticipate a multiband frequency coverage of at least 512\,GHz and significantly wider instantaneous bandwidth in each band than the current 7.5-GHz bandwidth. This can further improve the detection sensitivity of pulsars and transients, but most importantly, offers an extremely powerful mean for spectral studies of these objects. 

\section{Synergies with current and future facilities}

Coordinated observations with telescopes operating at lower frequencies and in the future, the SKA, will be essential for studying the emission of pulsar, magnetar and fast transient. Broad frequency coverage connecting the GHz and mm regimes is required to characterize spectral turnovers or cutoffs \citep{Philippov2020}, flat or inverted spectral components, and any high-frequency emission predicted by incoherent or inverse-Compton mechanisms \citep{mic82,Crusius-Watzel2001,Lyutikov2013}. For pulsars, emission at high and low frequencies are expected to trace different emission heights and combining both is necessary to establish a complete picture of the emission geometry in the magnetosphere \citep{naa+11}. In particular for fast transients and magnetars, whose emission is highly time-variable, low-frequency observations provide the detection rates needed to trigger prompt follow-up observations at mm-wavelengths during their radio-active windows.

Synergy between ALMA2040 and other facilities is also highly important in the context of GC. Propagation effects toward the GC, such as scattering, dispersion, and Faraday rotation, originate in the dense, magnetized interstellar medium (ISM) of the central few hundred parsecs. The extreme and variable magneto-ionic environment revealed by the GC magnetar PSR~J1745$-$2900 illustrates how strongly these effects are tied to the local ISM \citep{Desvignes2018}. While these effects are most prominent at low frequencies, they may also be physically associated with the molecular gas and dust that high-resolution ALMA2040 images will reveal. Together, these complementary observations can probe the ISM and magnetic environment of the GC from multiple perspectives. 

For pulsar searching in the GC, low-frequency surveys, such as SKA-mid GC surveys, would be the most sensitive to MSPs in the weak-scattering regime \citep{2025OJAp....854252A}. However, in the strong-scattering case, observations at \(\gtrsim 40\) GHz are required, and only ALMA2040 offers both the sensitivity and frequency coverage for this task. Meanwhile, timing a pulsar in the vicinity of \sgra\ can be affected by perturbations from other stellar objects in the region, which could be comparable to or significantly larger than the relativistic effects in the timing data \citep{lwk12,pwk16}. Thus, other means of understanding stellar motions, particularly in the core of the GC, would be valuable for assessing and modeling the impact of stellar perturbations on the pulsar's orbit \citep{GRAVITY:2019}.

\section*{Acknowledgements}
We thank Alejandro Mus Mejías and Wu Jiang for providing valuable suggestions on writing this manuscript. This work is funded by the Chinese Academy of Sciences President’s International Fellowship Initiative (Grant No. 2026PVC0050). This work was developed as part of the broader ALMA2040 community effort. We acknowledge the Lorentz Center and NOVA for supporting the community workshops and coordination activities that helped initiate and shape the ALMA2040 White Paper process. LS was supported by the National SKA Program of China (2020SKA0120300), and the Max Planck Partner Group Program funded by the Max Planck Society.\\

\bibliographystyle{apj_mod}
\bibliography{journals,psrrefs-p1,psrrefs-p2,modrefs-p1,modrefs-p2,2024,2025,refs}
\end{document}